\documentclass[showpacs,preprintnumbers,prb,twocolumn]{revtex4}
\usepackage{amsmath}
\usepackage{amssymb}
\usepackage{graphicx}
\usepackage{dcolumn}
\usepackage{bm}

\begin{document}

\title{Coupled layered superconductor as a system of 2D Coulomb particles of two kinds.}
\author{A.~N.~Artemov}
\affiliation{Donetsk Institute for Physics and Technology,
National Academy of Sciences of Ukraine, 83114 Donetsk, Ukraine}
\email{artemov@kinetic.ac.donetsk.ua}
\date{\today}

\begin{abstract}
It is shown that the Josephson subsystem of the Lawrence-Doniach
model of layered superconductors in the London approximation can
be presented as a system with variable number of classical Coulomb
particles. This allows us to consider the vortex system of a
coupled layered superconductor as the system of these particles
and 2D-vortices interacting with each other. The grand partition
function of the system was written and transformed into the form
of field one. Thermodynamical properties of the model obtained was
studied. It is found that there is no a phase transition in the
system. Instead of this the model demonstrates the crossover from
a low temperature 3D behavior to high temperature 2D one which can
look as a phase transition for experimental purposes.
\end{abstract}


\maketitle

\section{INTRODUCTION}
Behavior of the vortex system of coupled layered superconductors
(SC) is one of the most interesting and difficult problem in the
thermodynamics of quasi two dimensional (2D) systems. It
demonstrates dimensional crossover and other issues of
dimensionality which are of perpetual interest in statistical
mechanics.

The Berezinskii-Kosterlitz-Thouless (BKT) phase transition
\cite{ber,kt} in the 2D-vortex system was proposed to explain
results of the experiment on measuring of the current-voltage
characteristics of BISCO monocrystal\cite{arto}. Such a type of
transition takes place in 2D systems in which topological defects
can exist. They can be considered as 2D Coulomb particles whose
interaction energy logarithmically depends on a distance between
them. Layered SC's without Josephson coupling between layers are
2D systems. 2D-vortices are topological defects in these systems.
2D vortex-antivortex dipoles of a finite size can arise as thermal
excitations at any temperatures and they dissociate into a gas of
free vortices above BKT transition temperature $T>T_{KT}$.
Properties of these systems are investigated very
well\cite{minh1}.

But in real layered SC's layers are coupled due to the Josephson
effect. They can not be considered as 2D systems. Simplest thermal
excitations in coupled layered SC's arise in the form of two
2D-vortices of opposite orientation localized in the same layer
which magnetic flux is closed by two Josephson vortices or
fluxons. The interaction energy of 2D-vortices in this case
logarithmically depends on a distance for a small dipole size and
asymptotically tends to linear dependence for a large
size\cite{feig}. Investigation of such systems is much more
difficult and less developed.

One of the way which was used to analyze the problem consists in
studying the thermodynamic of the system of classical particles
with the pair interaction potential containing both logarithmic
and linear terms (see for example Refs.
[\onlinecite{fisher,pierson,friesen}]). This approach takes into
account the linear tension of Josephson vortices as a portion of a
two-particle interaction energy only while to describe the problem
correctly it is necessary to take into consideration a Josephson
vortex subsystem as an independent participant of thermodynamic
processes which has own configuration energy, entropy, self-energy
and interacts with the 2D vortex subsystem.

The other approach to the problem is to investigate the
anisotropic 3D XY-model. It was shown that excitations of the same
type as in coupled layered SC can exist in this
model\cite{cat,minh2,choi}. The anisotropic 3D XY-model was
studied numerically using Monte Carlo simulations\cite{minh2,gou}
and analytically\cite{pires,jens}. Results of this investigations
show that the model undergoes the phase transition of the BKT type
at temperature $T_c$ higher then $T_{KT}$ of a planar model and
coupling of adjacent layers vanishes at the same temperature.

The attempt to solve the same problem, which our work is devoted
to, was made in Ref. [\onlinecite{benfatto}]. The authors base
their approach on the quantum 1D sine-Gordon model and use the
connection of this model with the 2D Coulomb gas system. The
partition function of the model is represented in terms of a
functional integral over two non-commuting variables. The model
behavior is studied by means of a perturbative  renormalization
group (RG) approach. The RG recursion relations (RR) derived in
the work have not a fixed point. However, authors analyzing RR
features come to conclusion that the system undergoes phase
transition which temperature $T_d$ is controlled by 2D-vortex core
energy.

The other approach to the problem discussed was proposed by
Pierson and Valls\cite{pierson2}. The authors considered the model
which they called a model of $XY$ layers with Lawrence-Doniach
(L-D) type interlayer couplings. It consists of 2D sine-Gordon
models in each layers and additional cosine terms with difference
of phases of adjacent layers, which describe the Josephson
coupling. To study the model behavior authors used RG approach
previously expanding the additional cosine term and keeping the
quadratic summand only. The RG RR's have not a fixed point at a
small value of the coupling constant but authors give arguments
for existing of that at a larger value beyond the range of
validity of RR's.

From our point of view such an approach to analyze the model is
inadequate to the problem discussed. As N\'{a}ndori et
al.\cite{nandori} showed in the case of the quadratic interlayer
coupling the model is equivalent to the system of classical
particles with long-range pair interaction. So, such an approach
excludes from consideration the contribution of the entropy of the
Josephson subsystem and the linear tension in the 2D-vortex
interaction energy.

In this work we show that the model considered by Pierson and
Valls\cite{pierson2} follows from the L-D model\cite{ld} in the
London approximation and that its properties is determined by
behavior of order parameter singular points of two kinds. They can
be considered as classical Coulomb particles which can be
associated with 2D-vortices and the Josephson subsystem.

The paper is organized as follow. In Sec. II we formulate the L-D
model in the London approximation which describes the system of
Josephson vortices and transform it into the form available for
further generalization. In Sec. III we introduce 2D-vortices in
the L-D model and show the close analogy between the 2D-vortex and
the Josephson systems. In Sec. IV the model of the vortex system
of a coupled layered SC is constructed as a generalization of
models considered in Sec. II and III. The behavior of the model is
analyzed in Sec. V and VI by means of the RG approach and in the
mean field (MF) approximation. Concluding remarks are presented in
Sec. VII.

\section{Fluxon system in the Lawrence-Doniach model.}

\subsection{Model transformation.}

We start from the L-D model\cite{ld} in the London approximation
in which fluctuations of the order parameter modulus responsible
for transition in a superconducting state are neglected. In this
case the order parameter is normalized so that its modulus is unit
and does not depend on coordinates. Such a model describes only
the system of Josephson vortices in a layered superconductor
because 2D-vortices are associated with singular points in which
the order parameter turns into zero. In this section we transform
the model into the form suitable for its generalization and
discuss briefly its thermodynamic properties.

The Hamiltonian is the functional of the phase $\theta$ of the
order parameter and the vector-potential $\textbf{A}$ of magnetic
field
\begin{widetext}
\begin{equation}\label{e2.1}
H=\frac{\phi_0^2}{16\pi^3\Lambda(T)}\sum_n\int d\textbf{r}dz
\left[\left(\nabla\theta_n(\textbf{r})-\frac{2\pi}{\phi_0}\textbf{A}(\textbf{r},z)\right)^2-\frac{2}{\lambda_J^2}
\cos(\Omega_n(\textbf{r}))\right]\delta(z-ns) + \int
d\textbf{r}dz\frac{[\nabla\times
\textbf{A}(\textbf{r},z)]^2}{8\pi}.
\end{equation}
\end{widetext}
Here $\phi_0$ is the magnetic flux quantum,
$\Lambda(T)=2\lambda_{ab}(T)/s$,
$\lambda_J=s\lambda_c/\lambda_{ab}$, $\lambda_{ab}(T)$ and
$\lambda_c(T)$ is the London penetration depths parallel and
perpendicular to layers, $s$ is the period of layered structure
and
$\Omega_n(\textbf{r})=\theta_{n+1}(\textbf{r})-\theta_{n}(\textbf{r})
-(2\pi/\phi_0)\int_{ns}^{(n+1)s}dzA_z(\textbf{r},z)$ is the gauge
invariant phase difference. The partition function of the model
can be written in terms of a functional integral as
\begin{equation}\label{e2.2}
    Z=\int D\theta
    D\textbf{A}\exp\left\{-\beta H[\theta,\textbf{A}]\right\},
\end{equation}
where $\beta=1/T$.  The vector potential $\textbf{A}$ in this
expression is a Gaussian variable an can be integrated out. The
variable $\textbf{A}$ is a gauge field and this must be taken into
account on calculating. In the case considered it is convenient to
choose the gauge condition $A_z=0$. Resulting partition function
is
\begin{eqnarray}\label{e2.3.1}
\nonumber
  Z&=&\int D\theta\exp\left\{-\frac{1}{2J}\sum_{n,n'}\int d\textbf{r}
  \frac{\nabla\theta_n(\textbf{r})
  K_{n,n'}\nabla\theta_{n'}(\textbf{r})}{2\pi} \right. \\
    &+&\left. 2y_f\sum_n\int\frac{d\textbf{r}}{a^2}
  \cos\left(\theta_{n+1}(\textbf{r})-\theta_n(\textbf{r})\right)\right\}
\end{eqnarray}
where $J=4\pi^2\Lambda T/\phi_0^2$, $y_f$ is the fluxon fugacity
which is proportional to Josephson critical current density  and
the operator
$$
K_{n,n'}=\delta_{n,n'}-\int_{-\pi}^{\pi}\frac{dk}{2\pi}
\frac{\cos\left(k(n-n')\right)}{1+\frac{\Lambda}{s}k^2}
$$
takes into account magnetic interaction between layers. In systems
we are interested the condition $s\ll\lambda_{ab}$ is fulfilled
and the magnetic interaction leads to small corrections of order
$s^2/\lambda^2$ and does not change results qualitatively.
Therefore for the sake of simplicity we will omit all over the
paper these corrections and will take $K_{nn'}=\delta_{nn'}$. For
convenience of further references we write the partition function
of this reduced model because it is the initial point of our
constructions
\begin{eqnarray}\label{e2.3.2}\nonumber
  Z&=&\int D\theta\exp\left\{-\frac{1}{2J}\sum_{n}\int d\textbf{r}
  \frac{(\nabla\theta_n(\textbf{r}))^2}{2\pi}\right. \\
  & +& \left. 2y_f\sum_n\int\frac{d\textbf{r}}{a^2}
  \cos\left(\theta_{n+1}(\textbf{r})-\theta_n(\textbf{r})\right)\right\}
\end{eqnarray}

The field partition function (\ref{e2.3.2}) can be transformed
into the form of that of a gas of classical Coulomb particles. In
order to perform such a mapping we follow to the scenario used by
Nandori et al.\cite{nandori}. With this aim in view we carry out
the change of the variable $\theta=\sqrt{J}\varphi$ and present
the exponential function with the cosine term in its argument as a
product of two exponents
\begin{widetext}
$$
\exp\left\{2y_f\sum_n\int \frac{d\textbf{r}}{a^2}
\cos\left[\sqrt{J}(\varphi_{n+1}-\varphi_n)\right]\right\}
=\prod_n \exp\left\{y_f\int\frac{d\textbf{r}}{a^2}
e^{\imath\sqrt{J}(\varphi_{n+1}-\varphi_n)}\right\}
\exp\left\{y_f\int\frac{d\textbf{r}}{a^2}
e^{-\imath\sqrt{J}(\varphi_{n+1}-\varphi_n)}\right\}.
$$
Then we expand them into Tailor series. In the result after
obvious change of summation we obtain the expression required
\begin{eqnarray}\label{e2.4}
 \nonumber
  Z&=&\prod_n\left[\sum_{M_{n\pm}}\frac{1}{M_{n+}!}\frac{1}{M_{n-}!}
\left(e^{\beta\mu_f} \int\frac{d\textbf{r}_{j_n}}{\xi^2}
 \right)^{M_{n+}+M_{n-}} \right] \\
  &\times & \int D\varphi\exp \left\{-\frac{1}{2}\sum_{n}\int d\textbf{r}
  \frac{(\nabla\varphi_n(\textbf{r}))^2}{2\pi}
   +\imath\sum_n\sum_{j_n=1}^{M_{n+}+M_{n-}}q_{j_n}
   \left(\varphi_{n+1}(\textbf{r}_{j_n})-\varphi_n(\textbf{r}_{j_n})
   \right)\right\}.
\end{eqnarray}
\end{widetext}
In this expression $q_{j_n}=\pm\sqrt{J}$ are charges of the fluxon
particles, $M_{n+}$($M_{n-}$) are numbers of particles with a
positive (negative) charge in $n$th layer, $\mu_f$ is the chemical
potential of the particles and length $a$ is replaced by the
correlation length $\xi$.

The particles under consideration, apparently, are not real ones.
They look as dipoles which poles have the same in-plain positions
but are located in adjacent layers and can be presented as a point
current source and sink. In fact, such a particle is a piece of a
current line connecting adjacent layers.

Further transformation of the partition function is possible if we
notice that $\varphi$ is the Gaussian variable and can be
integrated out. Such a procedure leads to the partition function
of the system of classical particles with a pair interaction
\begin{eqnarray}\label{e2.5}
  Z &=&\prod_n\left[\sum_{M_{n\pm}}\frac{1}{M_{n+}!}\frac{1}{M_{n-}!}
\left(e^{\beta\mu_f} \int\frac{d\textbf{r}_{j_n}}{\xi^2}
 \right)^{M_{n+}+M_{n-}} \right]\nonumber \\
  &&\!\!\!\times \exp\left\{-\frac{1}{2}\sum_{n,n'}\sum_{j_n,j_{n'}}q_{j_n}q_{j_{n'}}
   v_{nn'}(|\textbf{r}_{j_n}-\textbf{r}_{j_{n'}} |)\right\}.
\end{eqnarray}

The potential $v_{nn'}$ corresponds to 2D Coulomb interaction of
above-mentioned fluxon particles
\begin{eqnarray}\label{e2.6}\nonumber
v_{nn'}(\textbf{r}_{j_n}-\textbf{r}_{j_{n'}})&& \\
=\ln\frac{|\textbf{r}_{j_n}-\textbf{r}_{j_{n'}} |}{\xi}
&&\!\!\!\!\!\left(\delta_{n,n'}-\frac{1}{2}\delta_{n,n'-1}-
\frac{1}{2}\delta_{n,n'+1}\right).
\end{eqnarray}

There is another way to obtain the partition function
(\ref{e2.5}). We can start from Eq. (\ref{e2.3.2}) with $y_f=0$
and assume that the order parameter possesses singular points in
which the conditions
\begin{eqnarray}\label{e2.7}\nonumber
 (\nabla,\nabla\theta_{n'}(\textbf{r}-\textbf{r}_{j_n}))& =& \pm 2\pi
J\delta(\textbf{r}-\textbf{r}_{j_n})
\left(\delta_{n+1,n'}-\delta_{n,n'}\right), \\
\left[\nabla,\nabla\theta_n(\textbf{r})\right]&=&0
\end{eqnarray}
are satisfied. It is easy to show that the energy of interaction
of singular points defined by these conditions is the same Coulomb
potential (\ref{e2.6}) as that of particles discussed.

Thus, the field model (\ref{e2.3.2}) of fluxon subsystems can be
presented as the system with variable number of classical
particles with  pair interaction. Such a point of view on the
fluxon system will be very useful in Sec.IV to construct the model
in which fluxon and 2D-vortex subsystems are joined.

\subsection{Thermodynamic properties of the model.}
Peculiarities of behavior of the model will reveal themselves in
the properties of the joined model. To better understand them we
discuss briefly the thermodynamic properties of the fluxon
subsystem. The main interest will be directed to an order and
temperature of the phase transition. Similar model was studied by
Horovitz\cite{hor}. He showed that the second order phase
transition of the BKT type takes place in the system. Here we
discuss results of our examination of the model which are very
close to Horovitz ones.

\begin{figure}[h]
  \includegraphics[width=8.5cm]{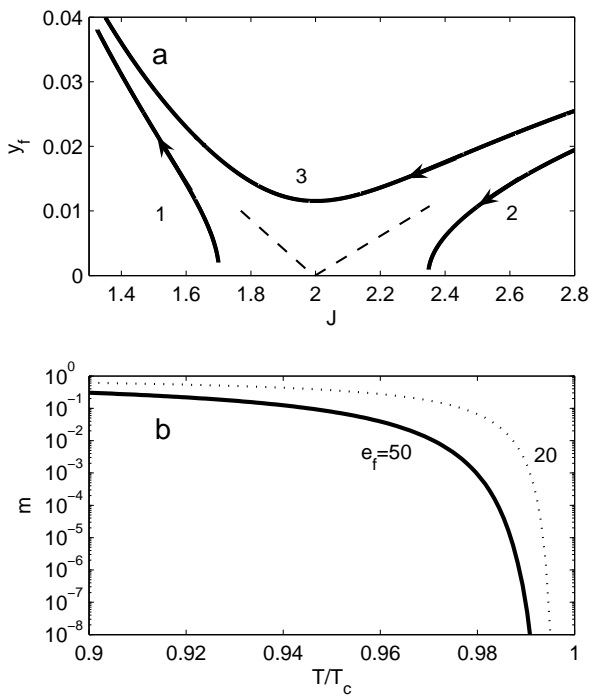}\\
  \caption{(a) Phase trajectories of the fluxon system RG equations
  in the vicinity of the fixed point ($J=2$, $y_f=0$) of saddle type.
  (b) Temperature dependence of the fluxon particle concentration
  for the systems with different anisotropy $e_f=50$ and $20$
  not far from the phase transition } \label{fig1}
\end{figure}

We perform the momentum space RG study of Eq.(\ref{e2.3.2}). RR's
are a the set of two equations in parameters of the initial
Hamiltonian
\begin{eqnarray} \label{e2.8.1}
 %\nonumber
  \frac{dy_f}{d\tau} &=& \left(2-J\right)y_f, \\
   \label{e2.8.2}
  \frac{d J}{d\tau} &=& -8\pi^2J^3y_f^2.
\end{eqnarray}
The phase portrait of the set is plotted in Fig. 1(a). It is the
conventional picture of flows in the vicinity of a fixed point of
the saddle type which is defined by the conditions ($J=2$,
$y_f=0$). Existence of the fixed point means that a second order
phase transition takes place in the system at the temperature
defined by the expression
\begin{equation}\label{e2.9}
    T_f=\frac{\phi_0^2}{2\pi^2\Lambda(T_f)}
\end{equation}
which follows from the condition $J=2$. This temperature is placed
above than that of the BKT transition $T_{KT}$ in the 2D-vortex
system of decoupled layered SC and is much more close to the
temperature $T_c$ of the transition of a sample into a
superconducting state. Another feature of the transition is the
directions of the RG flows. They are shown by arrows in Fig.
\ref{fig1}(a) and, as it will be shown in next section (see Fig.
 \ref{fig2}(a)), are opposite to those of 2D-vortex system.

The fugacity $y_f$ turns to zero in the fixed point. As $y_f$ is
proportional to Josephson critical current density this means that
it vanishes in the transition point $T=T_f$ too and
superconducting layers becomes decoupled at high temperatures
$T>T_f$.

Let us consider also the temperature dependence of the
concentration of fluxon particles. We found it in the MF
approximation. The free energy of the system per the unit area of
one layer was obtained from Eq.(\ref{e2.5}) by means of the
thermodynamic perturbation theory in the ring
approximation\cite{bal}. Under the suppositions that the system is
neutral, equilibrium numbers of particles doesn't depend on the
layer index ($M_{n+}=M_{n-}=M$) and  the screening length
$\delta_f\gg \xi$ the expression for the dimensionless free energy
takes the form
\begin{equation}\label{e2.10}
    f=\frac{\beta F\pi\xi^2}{S}=2m\left(\ln(m)
    -1\right)-Jm\ln(4Jm)+mJe_f.
\end{equation}
Here $m=\pi\xi^2M/S$ is the dimensionless particle concentration,
$S$ is the layer area, and $1/\sqrt{4Jm}=\delta_f/\xi$ is the
dimensionless Debye screening length. The last term is, in fact,
the chemical potential one. It controls the system anisotropy and
we consider $e_f$ as a phenomenological parameter. The free energy
($\ref{e2.10}$) reaches its minimal value at the particle
concentration
\begin{equation}\label{e2.11}
    m=\left(\frac{4J}{e^{e_f-1}}\right)^{\frac{J}{2-J}}.
\end{equation}
This solution is stable at $J<2$ ($T<T_f$). In the point $J=2$ it
loses the stability and at $J>2$ ($T>T_f$) equilibrium value of
the concentration is $m=0$. So, the MF approximation leads to the
same temperature (\ref{e2.9}) of the phase transition in
correspondence with RG results.

The temperature dependence of the equilibrium particle
concentration is plotted in Fig. 1(b) for systems with different
anisotropy. To derive this picture the model temperature
dependence of the London length $\lambda_{ab}(T)\sim
(1-T/T_c)^{-1/2}$ was used and parameters of the model were chosen
in such a way to fix $T_{KT}/T_c=0.98$. These options result in
$T_f/T_c\approx 0.998$. It is seen that the free particle
concentration $m$ in a more anisotropic system ($e_f=50$) always
is less then that in a less anisotropic ($e_f=20$). We can see
also some correlations between temperature dependencies of the
free fluxon particle concentration and critical density of the
Josephson current: both of them reach their maximum values at
$T=0$, both monotonically decrease when the temperature increases,
both are greater in less anisotropic systems, both turn into zero
at $T=T_f$ and vanish at $T>T_f$.

\section{2D-vortex system in the Lawrence-Doniach model.}

\subsection{Model constructing and transformation}

To describe the 2D-vortex system in the L-D model we use the
Eq.(\ref{e2.3.2}) with $y_f=0$. 2D-vortices are topological
excitations and can be introduced in the model as singular points
of the order parameter placed in positions $\textbf{r}_{j_n}$ of
the $n$th layer and defined by the conditions
\begin{equation}\label{e3.1}
\left[\nabla,\nabla\theta_n(\textbf{r})\right]=2\pi
s_{j_n}\textbf{z}\delta(\textbf{r}-\textbf{r}_{j_n}), \phantom{a}
\left(\nabla,\nabla\theta_n(\textbf{r})\right)=0.
\end{equation}
The symbol $s_{j_n}=\pm 1$ means the sign of a charge of the
vortex placed in the position $\textbf{r}_{j_n}$ of the $n$th
plane and $\textbf{z}$ is the unit vector in $z$ direction. The
energy of interaction of two vortices defined in such a way and
placed in the points $\textbf{r}_{i_n}$ and $\textbf{r}_{j_n}$ of
the same $n$th layer is
\begin{eqnarray}\label{e3.2} \nonumber
U(\textbf{r}_{i_n}-\textbf{r}_{j_n})&=&
-Q_{i_n}Q_{j_n}u(\textbf{r}_{i_n}-\textbf{r}_{j_n}) \\
=&-&Q_{i_n}Q_{j_n}\ln
\left(\frac{|\textbf{r}_{i_n}-\textbf{r}_{j_n}|}{\xi}\right) ,
\end{eqnarray}
where $Q_{j_n}=s_{j_n}/\sqrt{J}$ is the 2D-vortex charge. The
interaction potential (\ref{e3.2}) is the 2D Coulomb one and obeys
the 2D Poisson equation
\begin{equation}\label{e3.3}
    -\triangle u(\textbf{r}) = 2\pi\delta(\textbf{r}).
\end{equation}

One can write the partition function of the system of 2D-vortices
considering them as classical massless particles with the
interaction potential (\ref{e3.2})
\begin{widetext}
\begin{equation}\label{e3.4}
    Z=\prod_n\left[\sum_{N_{n\pm}}\frac{1}{N_{n+}!}\frac{1}{N_{n-}!}
    \left(e^{\beta\mu_v}\int\frac{d\textbf{r}_{j_n}}{\xi^2}
    \right)^{N_{n+}+N_{n-}}\right]
    \exp\left\{-\frac{1}{2}\sum_{n}\sum_{i_nj_n} Q_{i_n}Q_{j_{n}}
    u(\textbf{r}_{i_n}-\textbf{r}_{j_{n}})\right\}.
\end{equation}
\end{widetext}
Here $N_{n\pm}$ are the numbers of particles with the positive and
negative sign of charge and $\mu_v$ is the 2D-vortex chemical
potential. This partition function can be transformed into the
forms of that of a system of particles into self-consistent field
and that of a field system\cite{art2}. With this aim in view it is
need to multiply the exponential function in Eq.(\ref{e3.4}) by
the unit, which is represented as the ratio of two identical
Gaussian integrals over a field variable $\varphi$
\begin{widetext}
\begin{equation}\label{e3.5}
  \exp\left\{-\frac{1}{2}\sum_{n}\sum_{i_nj_{n}} Q_{i_n}Q_{j_{n}}
    u_{n}(\textbf{r}_{i_n}-\textbf{r}_{j_{n}})\right\}
    \frac{1}{Z_0}\int D\varphi \exp\left\{-\frac{1}{2}\sum_{n}\int d\textbf{r}
    \frac{(\nabla\varphi_n)^2}{2\pi} \right\}.
\end{equation}
\end{widetext}
The integral in the denominator is just a
normalizing constant $Z_0$. The next step is to change of the
variable in the numerator
$$
\varphi_n(\textbf{r})=\varphi_n(\textbf{r})+\imath
\sum_{j_n}Q_{j_n}u_{n}(\textbf{r}-\textbf{r}_{j_n})
$$
which is chosen in such a way that after integrating by parts over
\textbf{r} and using Eq.(\ref{e3.3}) it leads to compensation of
the sum of interaction potentials in the power of exponential
function (\ref{e3.4}). As a result, the partition function
(\ref{e3.4}) can be rewritten in the form of that of a system of
particles in self-consistent field
\begin{widetext}
\begin{equation}\label{e3.6}
    Z=\prod_n\left[\sum_{N_{n\pm}}\frac{1}{N_{n+}!}\frac{1}{N_{n-}!}
    \left(e^{\beta\mu_v}\int\frac{d\textbf{r}_{j_n}}{\xi^2}
    \right)^{N_{n+}+N_{n-}}\right]
   \int D\varphi\exp\left\{-\frac{1}{2}\sum_{n}\int d\textbf{r}
    \frac{(\nabla\varphi_n)^2}{2\pi} +
    \imath\sum_{n}\sum_{j_n}Q_{j_n}\varphi_{n}(\textbf{r}_{j_n}) \right\}
\end{equation}
%\end{widetext}
In the Eq.(\ref{e3.6}) we can take the summation over particle
numbers $N_{n+}$ and $N_{n-}$ to obtain the field sine-Gordon
model. The substitution of the variable in the functional
integration $\varphi=\theta/\sqrt{J}$ returns us to the initial
variable $\theta$, which is the phase of the order parameter. The
partition function finally takes the form
%\begin{widetext}
\begin{eqnarray}\label{e3.7} \nonumber
    Z=\int D\theta\exp\left\{-\frac{1}{2J}\sum_{n}\int d\textbf{r}
    \frac{(\nabla\theta_n)^2}{2\pi}
     +2y_v\sum_{n}\int\frac{d\textbf{r}}{a^2}
    \cos\left(\frac{1}{J}\theta_n(\textbf{r})\right)\right\}.
\end{eqnarray}
\end{widetext}
Here $y_v=\exp\{\beta\mu_v\}$ is the 2D-vortex fugacity and
nonessential constant $Z_0$ is omitted.

Thus, the 2D-vortex system of a layered superconductor can be
represented in the same three forms as the fluxon one. Below we
discuss briefly thermodynamical properties of the 2D-vortex system
to compare them with those of the fluxon one.

\subsection{Thermodynamic properties of the model.}

Thermodynamical properties of this model was studied very much and
are well known\citep{minh1,hor}.

The momentum space RG approach based on Eq. (\ref{e3.7}) gives the
RR's
\begin{eqnarray}\label{e3.8.1}
\frac{dy_v}{d\tau}&=&\left(2-\frac{1}{2J}\right)y_v ,\\
\label{e3.8.2}
 \frac{d J}{d\tau}&=&\frac{4\pi^2}{J}y_v^2.
\end{eqnarray}
Existence of the fixed point ($J=1/4$, $y_v=0$) of the saddle type
means that the second order phase transition takes place in the
system at the temperature
\begin{equation}\label{e3.9}
    T_{KT}=\frac{\phi_0^2}{16\pi^2\Lambda(T_{KT})}.
\end{equation}
This is the conventional BKT transition. It is easy to see
comparing Eqs. (\ref{e3.9}) and (\ref{e2.9}) that $T_{KT}<T_f$.
The phase portrait of the Eqs. (\ref{e3.8.1}) and (\ref{e3.8.2})
is plotted in Fig. 2(a). Comparison of Figs. 1(a) and 2(a) shows
that RG flows of the fluxon and the 2D-vortex systems are
oppositely directed.

To find temperature dependence of the 2D-vortex concentration we
used the MF approximation. The free energy of the system as a
function of the vortex number can be obtained from the partition
function Eq. (\ref{e3.4}) by means of the thermodynamic
perturbation theory in the ring approximation \cite{bal}. Under
supposition that the system is neutral the dimensionless free
energy per the area unit $\pi\xi^2$ of one layer takes the form
\begin{equation}\label{e3.10}
    f=2n(\ln n-1)-\frac{n}{2J}\left(\ln \frac{4n}{J}-1\right)
    + n\frac{e_v}{2J}.
\end{equation}
Here $n=\pi\xi^2N/S$ is the dimensionless vortex concentration,
$\sqrt{J/4n}=\delta_v/\xi$ is the dimensionless Debye screening
length, $e_v/4J$ is the vortex core energy.
\begin{figure}[h]
  \includegraphics[width=8.5cm]{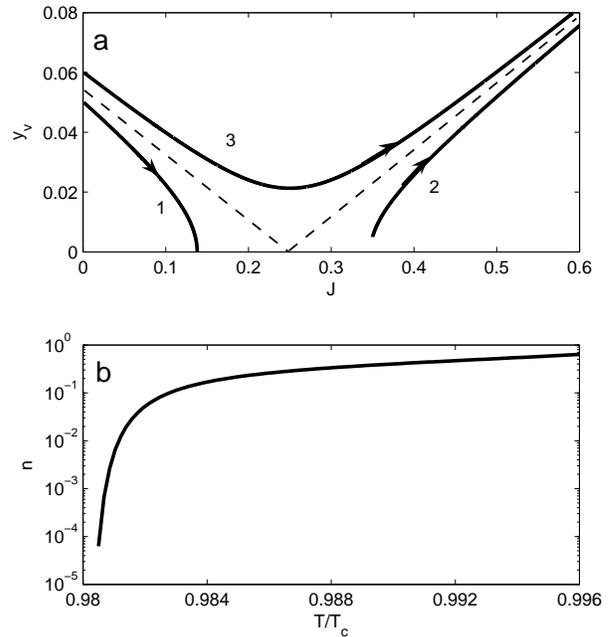}\\
  \caption{(a) Phase trajectories of the 2D vortex system RG equations
  in vicinity of the fixed point ($J=1/4$, $y_v=0$).
  (b) Temperature dependence of the free 2D-vortex concentration. }
  \label{fig2}
\end{figure}

The equilibrium vortex concentration obeys the equation
\begin{equation}\label{e3.11}
    n=\left(\frac{4}{J}\texttt{e}^{-e_v}\right)^{\frac{1}{4J-1}}.
\end{equation}
This solution is stable at $T>T_{KT}$. At $T=T_{KT}$ it becomes
zero and loses the stability at $T<T_{KT}$ where zero solution
becomes stable. Temperature dependence of the concentration is
shown in Fig. \ref{fig2}(b). To obtain this picture we used model
temperature dependence $\lambda_{ab}(T)\sim(1-T/T_c)^{-1/2}$ and
chouse model parameters to specify $T_{KT}/T_c=0.98$.

It is interesting to compare temperature dependencies of fluxon
particles Fig. \ref{fig1}(b) and 2D-vortices Fig. \ref{fig2}(b)
concentrations. A state with a finite concentration of 2D-vortices
is stable in the temperature range $T_{KT}<T<T_c$ which is located
higher the phase transition temperature while that of fluxon
particles is stable at lower temperatures $0<T<T_f$.

So, we considered two basic models. We found their common features
and revealed differences between them. By this we finish
consideration of the previous results. We have all we need to
construct the model of the vortex system of a coupled layered
superconductor. This is the subject of next section.

\section{Model of interacting 2D-vortex and fluxon system.}

Now as a result of previous consideration it is easy to see close
analogy of the fluxon and 2D-vortex subsystems of a layered SC.
Each of them can be presented either as a field model of a
sine-Gordon type or as a system of classical particles with a pair
Coulomb interaction. Difference between them consists in the
structure and the value of "charge" of the particles only. But our
aim is to develop the model describing a coupled layered SC which
contains both these subsystems. Therefore we will consider a
coupled layered SC as a system consisting of two kinds of 2D
Coulomb particles, namely, fluxon particles and 2D-vortices,
interacting with each other. Such a view on layered SC's allow us
to obtain the partition function of their vortex system to study
thermodynamic properties.

There are two ways to solve this problem. One can act in the
complete analogy with constructing of the 2D-vortex subsystem in
Sec. III taking into account singular points of the order
parameter defined by both conditions (\ref{e2.7}) and (\ref{e3.1})
to obtain 2D-vortex, fluxon particle and 2D-vortex - fluxon
particle pair interaction potentials and write a partition
function.

The other way is to construct the partition function of the system
of particles of both kinds in self-consistent field. Really,
existence of the interaction between the particles means that all
of them produce the same field and interact with it in accordance
with their structures and charges. In the analogy with Eqs.
(\ref{e2.4}) and (\ref{e3.6}) we can write
\begin{widetext}
\begin{eqnarray}\label{e4.1}
    Z&=&\prod_n\left[\sum_{N_{n\pm}}\frac{1}{N_{n+}!}\frac{1}{N_{n-}!}
    \left(e^{\beta\mu_v}\int\frac{d\textbf{r}_{j_n}}{\xi^2}
    \right)^{N_{n+}+N_{n-}}
    \sum_{M_{n\pm}}\frac{1}{M_{n+}!}\frac{1}{M_{n-}!}
\left(e^{\beta\mu_f} \int\frac{d\textbf{r}_{\alpha_n}}{\xi^2}
 \right)^{M_{n+}+M_{n-}}
    \right]\times \nonumber \\
    &&\int D\varphi\exp\left\{-\frac{1}{2}\sum_{n}\int d\textbf{r}
    \frac{(\nabla\varphi_n)^2}{2\pi} +
    \imath\sum_{n}\sum_{j_n}Q_{j_n}\varphi_{n}(\textbf{r}_{j_n})
     +\imath\sum_n\sum_{\alpha_n}q_{\alpha_n}
   \left(\varphi_{n+1}(\textbf{r}_{\alpha_n})-\varphi_n(\textbf{r}_{\alpha_n})
   \right)\right\}.
\end{eqnarray}
\end{widetext}

Next step is to carry out the sums over all numbers of particles.
These sums are easy to calculate because the result of integration
over particle coordinates $\textbf{r}_{j_n}$ and
$\textbf{r}_{\alpha_n}$ do not depend on particle indices and sums
are just Taylor series of exponential functions. As a result, the
partition function after change of a variable
$\varphi=\theta/\sqrt{J}$ takes the form
\begin{widetext}
\begin{equation}\label{e4.2}
    Z=\int D\theta \exp\left\{-\frac{1}{2J}\sum_n\int d\textbf{r}
    \frac{(\nabla\theta_n(\textbf{r}))^2}{2\pi}
    +2y_v\sum_n\int \frac{d\textbf{r}}{a^2}
    \cos\left(\frac{1}{J}\theta_n(\textbf{r})\right)
    +2y_f\sum_n\int \frac{d\textbf{r}}{a^2}
    \cos(\theta_{n+1}(\textbf{r})-\theta_{n}(\textbf{r}))\right\}.
\end{equation}
\end{widetext}

The expression obtained is the model we developed. It describes
the fluxon subsystem, the 2D-vortex subsystem as well as
interaction between them and is, in fact, the reduced L-D model in
which order parameter fluctuations is neglected. This model
includes the initial models, which was the base to develop the
model, as limiting cases. In the case $y_v=0$ the model
(\ref{e4.2}) reverts back to that of the fluxon system
(\ref{e2.3.2}) and the condition $y_f=0$ convert it into Eq.
(\ref{e3.7}) which is the partition function of the 2D-vortex
subsystem.

The model considered coincides actually with that proposed by
Pierson et al.\cite{pierson2}. To investigate the model authors
expanded the second cosine term and performed a RG analysis of the
reduced model. N\'{a}ndori et al.\cite{nandori} showed that the
reduced model is equivalent to a gas of topological defects with
long-range interaction potentials. But as it was shown in this
section the model \ref{e4.2} is equivalent to the gas of classical
Coulomb particles of two kinds. In contrast to work\cite{pierson2}
we study the model taking this into account.

\section{Renormalization grope analysis.}

We derived the perturbative RG equations in the parameters of the
Hamiltonian of the model (\ref{e4.2}) by means of the momentum
space approach. Recursion relations we obtained
\begin{eqnarray}\label{e5.1.1}
    \frac{d J}{d\tau}&=&\frac{4\pi^2}{J}y_v^2-8\pi^2J^3y_f^2,\\
    \label{e5.1.2}
    \frac{dy_v}{d\tau}&=&\left(2-\frac{1}{2J}\right)y_v, \\
    \label{e5.1.3}
    \frac{dy_f}{d\tau}&=&\left(2-J\right)y_f
\end{eqnarray}
takes into account renormalization of the parameters of the
initial Hamiltonian only. It is easy to see that the set of
equations inherits the structures of sets
(\ref{e2.8.1}-\ref{e2.8.2}) and (\ref{e3.8.1}-\ref{e3.8.2}) as
well as the model (\ref{e4.2}) inherits the structures of models
(\ref{e2.3.2}) and (\ref{e3.7}). Equations in fugacities $y_v$ and
$y_f$ are just identical and the right-hand side (r.h.s.) of Eq.
(\ref{e5.1.1}) is the sum of those of Eqs. (\ref{e2.8.1}) and
(\ref{e3.8.1}). Such a structure of the equations results in that
some details of the model behavior are similar to that of the
fluxon or 2D-vortex models but another ones are quite different.

Behavior of the set considered is much more complicated. Main
peculiar features of the set are that the number of independent
dynamical variables rises from 2 to 3 and the r.h.s. of Eq.
(\ref{e5.1.1}) is not sign-definite. In the result the recursion
relations (\ref{e5.1.1})-(\ref{e5.1.3}), contrary to that of the
fluxon system (\ref{e2.8.1})-(\ref{e2.8.2}) and the 2D-vortex one
(\ref{e3.8.1})-(\ref{e3.8.2}), have not a fixed point and thus the
system does not undergo a second order phase transition.

\begin{figure}[h]
  \includegraphics[width=8.5cm]{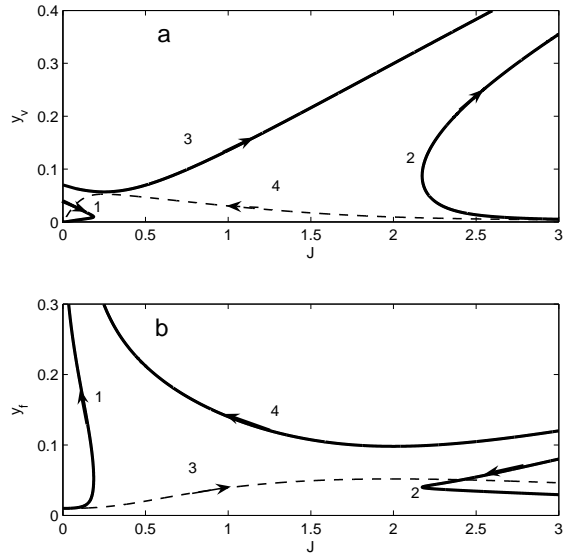}\\
  \caption{Projections of phase trajectories of the model RG equations.} \label{fig3}
\end{figure}

Typical trajectories of the RG recursion relations are shown in
Fig. \ref{fig3} as projections on planes $y_v-J$ and $y_f-J$. The
initial conditions are chosen in such a way to get the curves
which projections behavior, in a large extent, are similar to
those of corresponding trajectories of the fluxon and the
2D-vortex models. It is easy to see that the projection of
trajectory marked by 1 in Fig.~\ref{fig3}(a) at small values of
the scale variable $\tau$ behaves analogous to the trajectory 1 in
Fig. \ref{fig2}(a). But at some finite value of $\tau$ second term
in r.h.s. of Eq. (\ref{e5.1.1}) becomes dominated and the
derivative of $J$ changes the sign from plus to mines. It is a
turning point of the trajectory. After this point the analogy
mentioned disappears but instead analogy between the projection 1
in Fig. \ref{fig3}(b) and the trajectory 1 in Fig. \ref{fig1}(a)
appears. Similar analogies can be find between the projections of
trajectory 2 in Fig. \ref{fig3} and the trajectories 2 in Figs.
\ref{fig1}(a) and \ref{fig2}(a). There are else two analogies
between phase portraits of these systems. The projection 3 in Fig.
\ref{fig3}(a) is similar to the trajectory 3 in the phase portrait
of the RG equations of the 2D-vortex system, and the projection 4
in Fig. \ref{fig3}(b) behaves analogously to the trajectory 3 in
the phase portrait of the RG equations of the fluxon system.

Such an analogy between phase portraits of the model considered
and the models of independent 2D-vortex and fluxon systems allows
us to believe that there are some similarity between behavior of
these systems. But this analogy is observed in the space of
parameters of Hamiltonian and it does not allow us to draw a valid
conclusion about behavior of the vortex system in the space of
thermodynamical variables. Let us consider the phase portrait
shown in Fig. \ref{fig3} from another point of view.

The RG flow can be characterized, in addition to forms of
trajectories and direction of the system motion along them, also
by a rate of the system motion. If the recursion relations have a
fixed point the rate of motion along the trajectory is minimal in
the position which is nearest to this point and asymptotically
tends to zero when a trajectory approaches to it. Such a situation
takes place in the RG flows of the fluxon Fig. \ref{fig1}(a) and
the 2D-vortex Fig. \ref{fig2}(a) systems.

In the case of the model proposed the situation is different. The
RG flow Fig. \ref{fig3} does not have a fixed point. Instead, it
has a slow point in which a rate of motion along a trajectory is
lowest. This point can mimic a fixed point for experimental
purposes. The properties of models which demonstrate similar
behavior were investigated by Zumbach\cite{zumb}. He termed such a
situation as an "almost second order phase transition".

From such a point of view we can suppose that the model
asymptotically behaves almost the free fluxon and the 2D-vortex
systems in regions of low ($T<T_{KT}$) and high ($T>T_f$)
temperature and demonstrates the crossover from low temperature
$3D$ behavior to high temperature $2D$ one in vicinity of
temperature $T_s$ corresponding to a slow point. The turning
points of trajectories of different kinds (1-4 in Fig.\ref{fig3})
draw together in close vicinity of a slow point. This allow us to
evaluate the position of this point. Direct numerical calculation
of trajectories passing in the vicinity of slow point shows that
such a point corresponds to $J\approx 0.7$. The crossover
temperature which can be obtained from this relation obeys the
equation
\begin{equation}\label{e5.4}
    T_s=0.7\frac{\phi_0^2}{4\pi^2\Lambda(T_s)}.
\end{equation}

But a perturbative RG is not a very good approach to find a slow
point because the position of this point can be dependent on a
cut-off procedure. So, the expression (\ref{e5.4}) can be
considered as a rough evaluation of the crossover temperature
only. The fact in which we can be sure is that $T_{KT}<T_s<T_f$ or
the same $1/4<J_s<2$.

In completion of the section we note that a RG approach is the
examination of the hypothesis about a scaling invariance of the
model. In the case considered the RG does not have a fixed point
and, thus the hypothesis is not prove to be true, in contrast to
the independent 2D-vortex and fluxon models. In this situation we
have to use a different approach to understand reasons of the
scaling invariance breaking and to investigate the model behavior
in more details. This is a subject of the next section.

\section{Mean field analysis.}

\begin{figure}[h]
  \includegraphics[width=8.5cm]{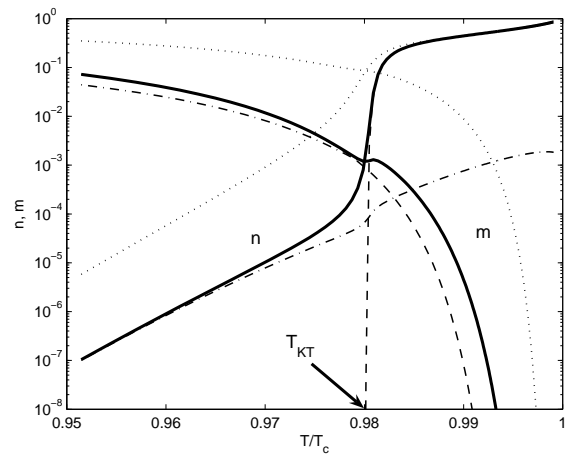}\\
  \caption{Solutions of equilibrium equations are shown by solid
  lines for the high anisotropic system $e_f=50$ and by dotted lines
  for the low anisotropic one $e_f=20$. Dash and dash and dot lines
  show asymptotical solutions Eqs. (\ref{e6.3}) and (\ref{e6.4})}.
  \label{fig4}
\end{figure}

In this section we will consider the model as a two subsystems of
classical particles with the long-range Coulomb interactions
(\ref{e2.6}) and (\ref{e3.2}) interacting with each other. To
obtain the free energy of the model in a MF approximation we use
the method developed in Ref. [\onlinecite{art1}]. This method
corresponds to the ring approximation in the case of the Coulomb
gas\cite{bal}. The system free energy obtained is a function of
concentrations of both 2D-vortices ($n$) and fluxon particles
($m$). Under the assumption that the system is neutral
($N_{n+}(M_{n+})=N_{n-}(M_{n-})=N(M)$) the free energy takes the
form
\begin{widetext}
\begin{equation}\label{e6.1}
    f=2n(\ln n -1)+2m(\ln m -1)
    -\left(\frac{n}{J}+2mJ\right)\ln\left(\sqrt{\frac{n}{J}} +
    \sqrt{\frac{n}{J}+4mJ}\right)
    +\frac{1}{2}\sqrt{\frac{n}{J}}\sqrt{\frac{n}{J}+4mJ}
    +n\frac{e_v}{2J}+mJe_f.
\end{equation}
%\end{widetext}
The condition of a minimum of the free energy is two
equations in equilibrium concentrations
\begin{eqnarray}
  &&2\ln n -\frac{1}{J}\ln\left(\sqrt{\frac{n}{J}}+
  \sqrt{\frac{n}{J}+4mJ}\right)+\frac{e_v}{2J}=0, \nonumber  \\
  &&2\ln m -2J\ln\left(\sqrt{\frac{n}{J}}+
  \sqrt{\frac{n}{J}+4mJ}\right) + J\frac{\sqrt{\frac{n}{J}}-
  \sqrt{\frac{n}{J}+4mJ}}{\sqrt{\frac{n}{J}}+
  \sqrt{\frac{n}{J}+4mJ}} +Je_f=0,
  \label{e6.2}
\end{eqnarray}
\end{widetext}
The solutions of the equations  obtained numerically are shown in
Fig.\ref{fig4} as temperature dependence of the equilibrium
concentrations of the 2D-vortices and fluxon particles by solid
lines for more anisotropic system ($e_f=50$) and by dotted lines
for less anisotropic ($e_f=20$) . These concentrations always are
above the corresponding values of the independent subsystems,
which are plotted by the dashed lines, and in contrast to those
are finite at any temperatures in the whole region $0<T<T_c$.

This result is easy to understand. There is BKT phase transition
in the independent systems of 2D-vortices and fluxon particles.
The mechanism realizing the transition is follow. A single
2D-vortex (fluxon particle) in an infinite sample has infinite
energy. As a result of this such a particle can not appear in the
system as a thermal fluctuation at low temperature $T<T_{KT}$
(high one $T>T_f$). But in a many body system the competition
between the configuration energy and entropy terms in the free
energy (see Eqs. (\ref{e2.10}) and (\ref{e3.10})) at $T>T_{KT}$
($T<T_f$) leads to instability of a zero concentration state and
to appearance of a finite concentration of free 2D-vortices
(fluxon particles). These particles differ from single ones
mentioned above because they have a finite energy due to Debye
screening. Such free vortices (fluxon particles) are, in fact,
quasi-particles and can appear in the system as thermal
fluctuations.

The model discussed describes systems of 2D-vortices and fluxon
particles interacting with each other. In the case of Coulomb
particles the interaction leads to their mutual screening. Thus, a
finite concentration of 2D-vortices (fluxon particles) at low
temperature $T<T_{KT}$ (high one $T>T_f$) exists due to screening
of them by fluxon particles (2D-vortices). This supposition is
easy to verify. The equilibrium equations (\ref{e6.2}) can be
simplified in the vicinity of $T=0$ ($T=T_c$) due to condition
$n\ll m$ ($m\ll n$). Asymptotical solutions of the equation take
the form
\begin{eqnarray}
\label{e6.3}\nonumber
  m &=&\left(\frac{4J}{e^{e_f-1}}\right)^{\frac{J}{2-J}},\\
  n &=& \exp\left\{\frac{1}{4J}\left(\ln 4Jm-e_v\right)\right\}
\end{eqnarray}
in the vicinity of $T=0$ and
\begin{eqnarray}
\label{e6.4}\nonumber
  n &=&\left(\frac{4}{J}\texttt{e}^{-e_v}\right)^{\frac{1}{4J-1}},\\
  m &=& \exp\left\{\frac{J}{2}\left(\ln \frac{4n}{J}-e_f\right)\right\}
\end{eqnarray}
close to $T=T_c$.

The solutions obtained make a clear sense. The concentration $m$
in (\ref{e6.3}) coincides with that in the independent fluxon
system (\ref{e2.11}). This is sequent of a very small
concentration $n$ of 2D-vortices at low temperatures $T\ll
T_{KT}$, which can be neglected in comparison with $m$. The
concentration of 2D-vortices $n$ in the limit considered is just
the Boltzmann expression $n=\exp\{-E_v/T\}$, where $E_v$ is the
vortex energy containing two terms. Second term is the vortex core
energy $e_v/4J$. First one is the energy of a 2D-vortex which
diverges logarithmically in the case of a single vortex but is
limited by the screening length $\delta_f$ due to a finite
concentration $m$ of fluxon particles, $4mJ=\xi^2/\delta_f^2$. The
Eq. (\ref{e6.4}) has an analogous structure: the 2D-vortex
concentration $n$ coincides with that in the independent system
Eq. (\ref{e3.11}) and the concentration $m$ is defined by the
self-energy of a fluxon particle screened by a finite
concentration of 2D-vortices, $4n/J=\xi^2/\delta_v^2$.

Asymptotical temperature dependencies of the concentrations of the
particles of both kinds are in a good agreement with the numerical
solution. The concentrations of 2D-vortices and fluxon particles
of independent systems are plotted in Fig. \ref{fig4} by dashed
lines. Asymptotical dependencies of concentrations of 2D-vortices
at low temperatures and fluxon particles at high temperatures are
shown by dash and dot lines.

It is seen that asymptotical behavior of the model at
$T\rightarrow 0$ and $T\rightarrow T_c$ is very close to that of
the independent 2D-vortex and fluxon systems. The concentrations
$n$ at $T\ll T_{KT}$ and $m$ at $T>T_f$ are exponentially small
but finite. This means that there is no a phase transition in the
model discussed but there is a crossover from 3D to 2D type of
behavior which takes place in the temperature interval
$T_{KT}<T<T_f$.

Thus results obtained in the framework of the MF approximation
agree quantitatively with that of the RG analysis. In terms of the
MF approximation they are conditioned by mutual screening of
particles of two kinds. The RG considers this as breaking of the
scaling invariance of the model. Now we can see that the reason
which leads to this is existence of two competitive lengthes,
namely, the screening lengthes $\delta_v$ and $\delta_f$.

\section{Conclusion}

We constructed and studied the model of the vortex system of
coupled layered superconductors which is based on the L-D one. The
main idea is to consider the Josephson and 2D-vortex subsystems as
the systems of singular points defined by the conditions
(\ref{e2.7}) and (\ref{e3.1}). Such an approach becomes possible
because both singularities interacts according to the 2D Coulomb
law. So, they can be interpreted as classical massless Coulomb
particles which are characterized by their charges and have
different structures. The model partition function can be
represented either as a grand partition function of the system
with a variable number of particles of two kinds (\ref{e4.1}) or
as a that of the field model (\ref{e4.2}).

The model was examined by means of the perturbative RG approach
and the MF approximation. Results obtained by both methods agree
qualitatively with each other. Both approaches show that there is
no phase transition in the model in the whole temperature interval
$0<T<T_c$ were the model is defined. But the model behaves
asymptotically at $T\rightarrow 0$ and $T\rightarrow T_c$ as
independent 2D-vortex and Josephson systems. Crossover from
low-temperature 3D behavior to high-temperature 2D one takes place
in the interval $T_{KT}<T<T_f$.

The effect of anisotropy can be cleared by means of comparing of
temperature dependencies of free particles concentrations of
systems with different value of the parameter $e_f$ in Fig.
\ref{fig4}. One can see that more anisotropic system ($e_f=50$)
behaves closer to that of independent 2D-vortex and fluxon
particle systems than less anisotropic ($e_f=20$). If the layered
system is anisotropic enough it can mimic the BKT phase transition
for experimental purposes.

\section*{Acknowledgments}

The author is thankful to A. E. Filippov and L. Benfatto for
useful personal discussions.

\end{document}